\documentclass[12pt]{iopart}
\usepackage{graphicx}
\usepackage[OT4]{fontenc}
\usepackage{bm}
\usepackage{dcolumn}
\usepackage{hyperref}
\hypersetup{
    colorlinks=false,
    pdfborder={0 0 0},
}
\usepackage{subfigure}
\usepackage{xspace}
\usepackage{upgreek}
\usepackage{ifthen}

\expandafter\let\csname equation*\endcsname\relax 
\expandafter\let\csname endequation*\endcsname\relax 
\usepackage{amsmath}
\usepackage{amssymb}

\newcommand{\mode}{reprint} 

\newcommand{\sizes}[2]{\ifthenelse{\equals{\mode}{reprint}}{#1}{#2}}

\newcommand{\gpe}{Gross-Pitaevskii equation\xspace}
\newcommand{\bec}{Bose-Einstein condensate\xspace}

\newcommand{\dw}{double-well\xspace}
\newcommand{\tw}{triple-well\xspace}
\newcommand{\bh}{Bose-Hubbard\xspace}

\newcommand{\qu}[1]{``#1''}

\newcommand{\bb}[1]{\left(#1\right)}                                		
\newcommand{\absv}[1]{\left|#1\right|}                             			
\newcommand{\absvsq}[1]{\absv{#1}^2}	                            		
\newcommand{\ket}[1]{\left|#1\right\rangle}  							 	
\newcommand{\unit}[1]{\ensuremath{\, \mathrm{#1}}}                     		

\renewcommand{\vec}[1]{\bi{#1}}                             
\newcommand{\vecr}{\ensuremath{\vec{r}}\xspace}                     

\newcommand{\partiald}[1]{\frac{\partial}{\partial #1}}             
\newcommand{\laplace}{\operatorname{\nabla^2}}                        

\newcommand{\integral}[1]{\int \! \mathrm{d} #1\,}                    
\newcommand{\intvol}{\integral{^3r}}								
\newcommand{\intvolp}{\integral{^3r'}}								
\newcommand{\intvold}{\intvol\!\intvolp}								

\newcommand{\expf}[1]{\ensuremath{\exp\bb{#1}}\xspace}	    		

\newcommand{\hamil}{\ensuremath{\operatorname{{\hat{H}}}}\xspace} 

\newcommand{\aop}{\ensuremath{\hat{a}^{\phantom\dagger}}\xspace} 
\newcommand{\aopd}{\ensuremath{\hat{a}^\dagger}\xspace} 
\newcommand{\nop}{\ensuremath{\hat{n}}\xspace} 

\newcommand{\add}{\ensuremath{a_\text{dd}}\xspace}                               
\newcommand{\aho}{\ensuremath{a_\text{ho}}\xspace} 
\newcommand{\acrit}{\ensuremath{a_\text{crit}}\xspace}                               
\newcommand{\Vdd}{\ensuremath{V_\text{dd}}\xspace}                               
\newcommand{\Cdd}{\ensuremath{C_\text{dd}}\xspace}                               
\newcommand{\Phidd}{\ensuremath{\Phi_\text{dd}}\xspace}                          
\newcommand{\Eint}{\ensuremath{E_\text{int}}\xspace}                             

\newcommand{\PsiOp}{\ensuremath{\operatorname{\hat{\Psi}}}\xspace}

\newcommand{\igopt}[3]{\ifthenelse{\equal{\mode}{reprint}}{\includegraphics[#1]{#3.eps}}{\includegraphics[#2]{#3.eps}}}
\newcommand{\ig}[3]{\igopt{width=#1\textwidth}{width=#2\textwidth}{#3}}
\newcommand{\figopt}[5]{
	\begin{figure}[#5]
	    \begin{center}
    		\ig{#1}{#2}{#3}
    	\end{center}
		\caption{#4}
		\label{fig:#3}
	\end{figure}
}

\begin{document}


\title{Mean-field description of dipolar bosons in triple-well potentials}

\date{\today}

\author{D Peter$^{1,2}$, K Paw\l owski$^{3,1}$, T Pfau$^{1}$ and K Rz\k{a}\.{z}ewski$^{3,4,1}$}
\address{$^{1}$ 5. Physikalisches Institut, Universit\"at Stuttgart, Pfaffenwaldring 57, 70550 Stuttgart, Germany}
\ead{peter@itp3.uni-stuttgart.de}

\address{$^{2}$ Institut f\"ur Theoretische Physik III, Universit\"at Stuttgart, Pfaffenwaldring 57, 70550 Stuttgart, Germany}

\address{$^{3}$ Center for Theoretical Physics, Polish Academy of Sciences, Al. Lotnik\'ow 32/46, 02-668 Warsaw, Poland}

\address{$^{4}$ Faculty of Mathematics and Sciences, Cardinal Stefan Wyszy\'nski University, ul. Dewajtis 5, 01-815, Warsaw, Poland}

\begin{abstract}
We investigate the ground state properties of a polarized dipolar \bec trapped in a triple-well potential. By solving the dipolar \gpe numerically for different geometries we identify states which reveal the non-local character of the interaction. Depending on the strength of the contact and dipolar interaction we depict the stable and unstable regions in parameter space.
\end{abstract}


\maketitle

\section{Introduction}
The physics of cold atoms in optical lattices is an active field of research in both experiment and theory~\cite{fisher1989,Jaksch1998,Greiner2002,greiner2002revival,Bloch2008}.
Interaction between the atoms and tunneling across the lattice 
are easily controlled using Feshbach resonances and by tuning the intensity of the external lasers, respectively. Already in one of the first experimental realizations of the system a quantum phase transition between a Mott insulator and a superfluid has been shown~\cite{Greiner2002}.
Together with the ultra-precise spatial resolution~\cite{Bakr2009,Sherson2010} this system is a good candidate for quantum simulators~\cite{Endres2011}
or, in the far future, even an element of a new generation of computers~\cite{calarco2004}.

The search for new phases is ongoing and dipolar interactions present additional possibilities~\cite{G'oral2002,Griesmaier2005,Beaufils2008}.
The major feature of the dipolar interaction is its long range character. 
Thus, the new phases are expected to reveal inter-site effects even in the case of suppressed tunneling. The first indication of such a phenomenon has already been shown in 
the dynamical properties of a Bose-Einstein condensation of very weakly interacting $^{39}$K~\cite{Fattori2008} and in the study of the stability of $^{52}$Cr, loaded in a 1D optical lattice~\cite{Mueller2011}.
Furthermore, the anisotropy of the dipolar interaction allows to control both inter- and on-site interactions by choosing the appropriate geometry of the lattice sites with respect to the polarization direction of the dipoles~\cite{Giovanazzi2002}. The easiest model systems consist of a few linked wells. In particular, the \dw system has received a lot of attention~\cite{Smerzi1997,Albiez2005}.
As the entanglement between the two macroscopically occupied modes has been demonstrated, it may be considered as an extension of a qubit~\cite{esteve2008,Riedel2010}. 
On the other hand, many interesting effects, including Josephson oscillations and quantum self trapping, were observed in the  frame of the mean field approximation~\cite{Zibold2010}.

In a recent discussion, the \tw potential, loaded with a dipolar gas, was studied~\cite{Lahaye2010}. 
An extended \bh model is used to describe the system, as in most related references concerning optical lattices~\cite{G'oral2002,Damski2003,Sengupta2005}. This model assumes fixed, occupation-independent parameters. For increasing particle numbers and interaction strength, this approximation is less reliable. 
Due to the interaction, the on-site spatial distribution of the atoms shrinks in the case of an attractive gas and broadens if the interactions are repulsive. 
The most dramatic case occurs for a dipolar gas when the number of atoms or the strength of interaction is above a critical value.
The sample collapses and then explodes in a so called Bose-Nova~\cite{Lahaye2008}.
As the shape of the atomic cloud changes with the interaction, the parameters of the \bh model cannot be uniquely defined.
In this paper we study the \tw case using a mean field approach.
Within this picture, the ground states for different geometries and interaction strength are discussed. We consider experimentally relevant parameters, especially for magnetic dipolar gases like Cr or Dy.

The paper is organized as follows. 
In \sref{sec:model} we describe both the \bh model and the mean field approach for a dipolar gas in an external \tw potential, which is modelled by overlapping Gaussian wells.
We discuss the limitations of the \bh approach and then switch to the mean field picture. An important new aspect arises. Depending on the geometry and the interaction parameters, the ground state solution may be unstable. In \sref{sec:results} we present a phase diagram for two specifically chosen geometries. The mean field results are compared to the ground states computed with the \bh model. The interesting states revealing the role of the inter-site effects are identified.

\section{System and model\label{sec:model}}
\begin{figure}[t]
    \begin{center}
	    \subfigure{
		    \ig{0.5}{0.5}{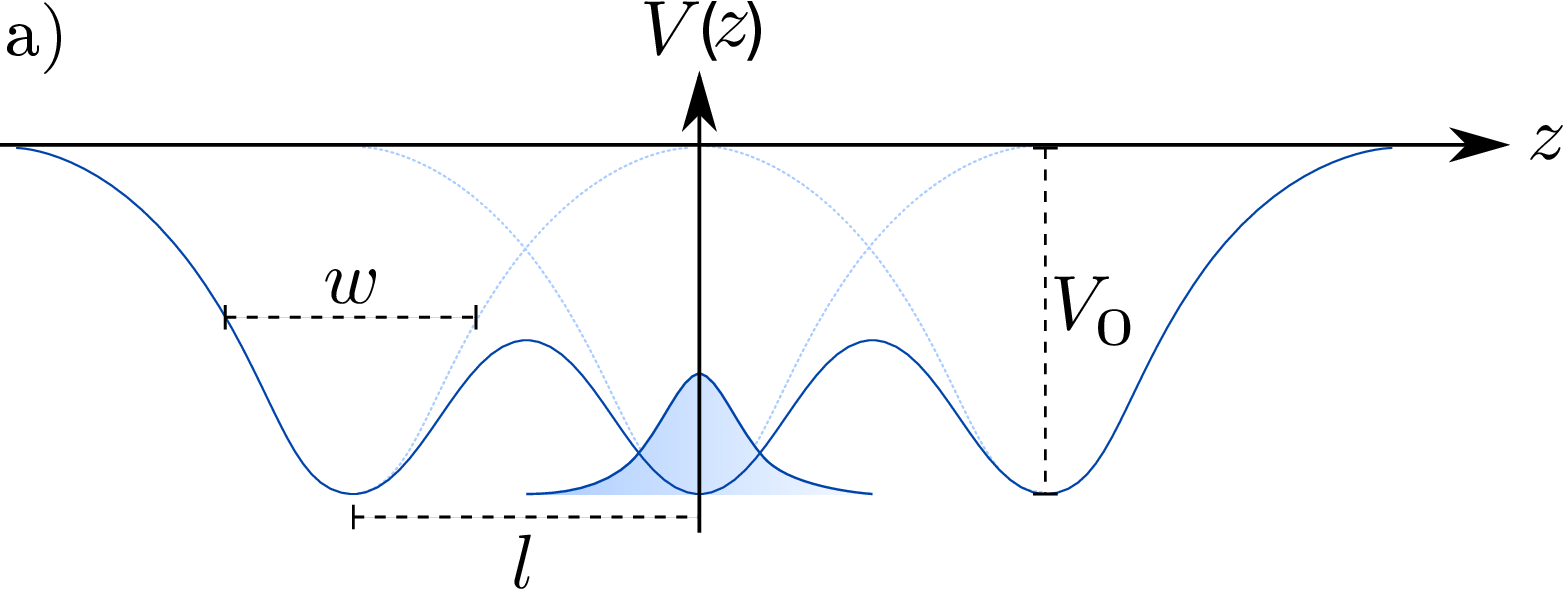}
		    \label{fig:triplewell}
	    }

	    \subfigure{
		    \ig{0.3}{0.35}{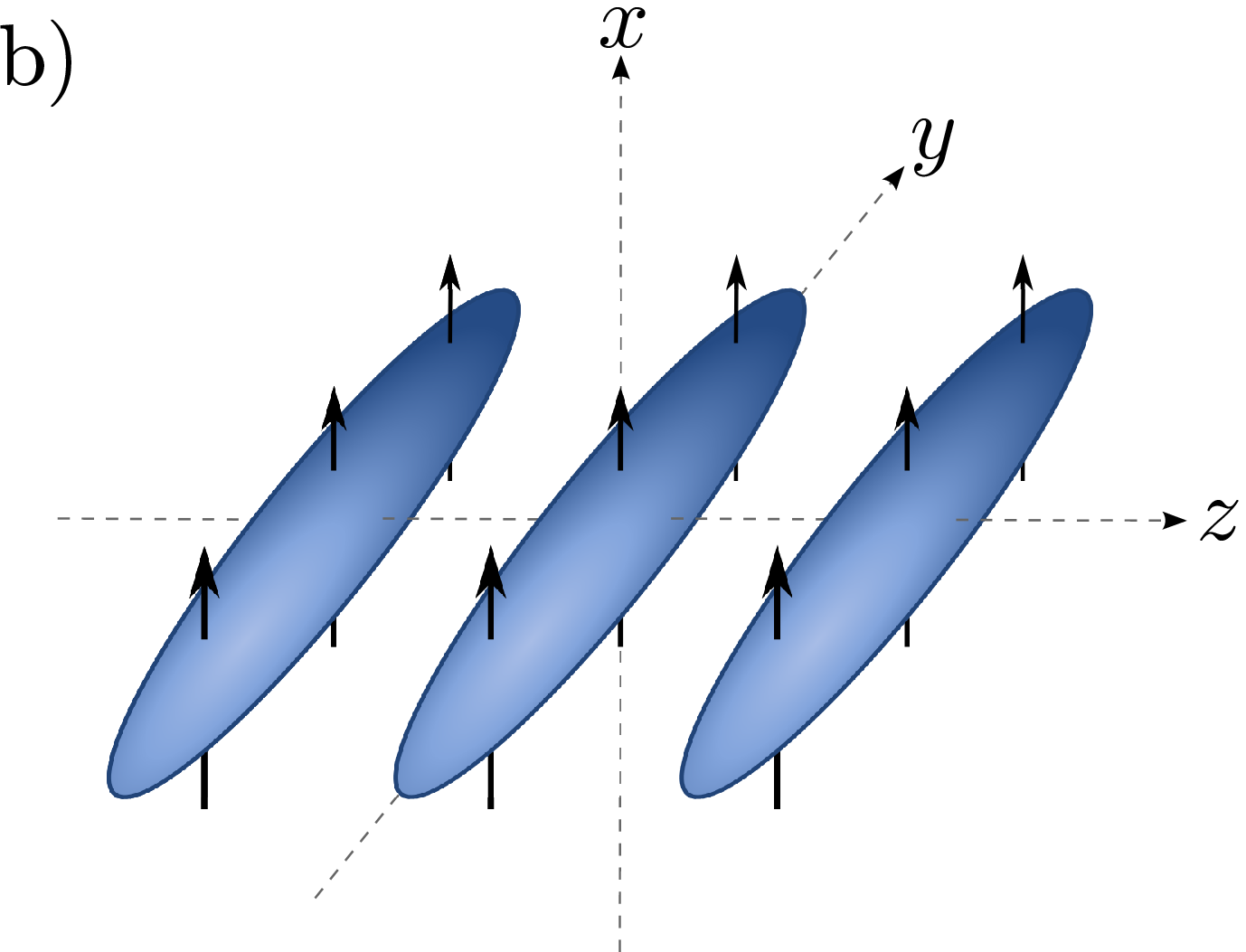}
		    \label{fig:geometry_rep}
	    }
	    \subfigure{
		    \ig{0.3}{0.35}{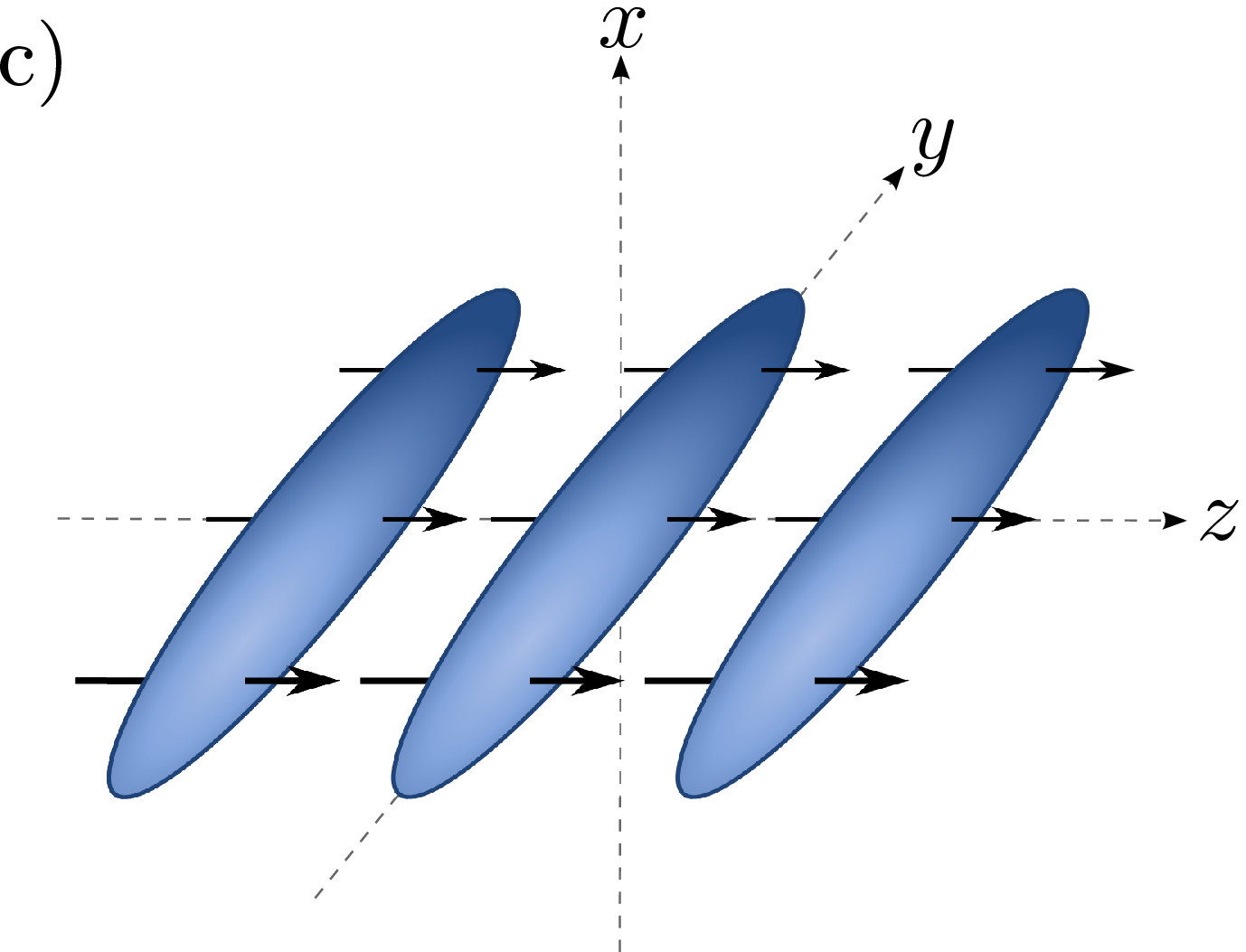}
		    \label{fig:geometry_attr}
	    }
	\end{center}
	\caption{(a) Cut of the \tw potential along symmetry axis. It is modelled by three overlapping Gaussians of width $w$.  (b) Repulsive and (c) attractive configuration of the dipoles. \label{fig:pot_geometry}}
\end{figure}

We consider a \bec consisting of $N$ dipolar particles with a dipole moment $\vec{d}$ which may be of either electric or magnetic origin. A strong external field is orienting the dipoles such that they all point in the same direction. The particles are subject to an external potential $V(\vecr)$ with three (nearly) equivalent minima, see \fref{fig:triplewell}, which is modelled by three overlapping Gaussian wells:
\begin{eqnarray}
    V(\vecr) = -V_0 \!\!\sum_{s=0,\pm 1}\!\! \expf{-\frac{2 x^2}{w_x^2} - \frac{2y^2}{w_y^2} - \frac{2(z\!-\!s\!\cdot\!l)^2}{w_z^2}}\!.
\end{eqnarray}
The parameter $V_0$ determines the centre depth of the
individual wells and the widths $w_i$ parametrise the geometry of
a single well (size in each direction). The spacing between
the wells is given by $l$.
A potential like this can be created by means of focused Gaussian laser beams~\cite{Lahaye2010}.

The particles are interacting via contact and dipolar interactions. The contact interaction is fully characterized by the scattering length $a$.
The dipolar interaction between polarized dipoles at positions $\vecr_1$, $\vecr_2$ is given by
\begin{eqnarray}
	\Vdd(\vec{r_1}, \vec{r_2}) =  \Cdd \, \frac{1 - 3 \cos^2(\vartheta)}{r^3}
\end{eqnarray}
where $r=|\vecr|=|\vecr_1-\vecr_2|$ is the inter-particle distance and $\vartheta$ is the angle between $\vecr$ and the dipole moment $\vec{d}$. The factor $\Cdd$ is equal to $d^2 \mu_0/4\pi$ for magnetic, and $d^2/4\pi \epsilon_0$ for electric dipoles. In analogy to the scattering length one introduces the length scale $\add = m \Cdd / 3\hbar^2$, characterizing the strength of the dipolar interaction~\cite{Koch2008}.

Throughout this work we are using a dimensionless system by measuring all lengths in units of the spacing~$l$, all energies in units of $\hbar^2/m l^2$ and time in units of $m l^2/\hbar$ ($m$ is the mass of the dipolar particles). We will keep the same notation for quantities with and without units, though.

\subsection{Extended \bh model} \label{sec:bhmodel}
Following the discussion in~\cite{Lahaye2010}, we outline how
a \bh model can be derived for the dipolar \tw system. First we assume that the three minima of the potential are well separated, such that the on-site wave functions $\phi_i(\vecr)$ for each site may be described by a single function: $\phi_i = \phi(\vecr - \vecr_i)$, with $\vecr_i$ the centre of the $i$-th well. The field operator $\PsiOp(\vecr) = \sum_{i=1}^3 \aop_i \phi_i(\vecr)$ can then be written in terms of the annihilation operators $\aop_i$ at site $i$. Interpreting these operators as annihilating a particle at site $i$ is correct as long as the potential is deep enough such that the overlap between the wave functions of two adjacent sites is small. The Hamiltonian is then expressed in \bh form as
\begin{eqnarray}
\hamil = -J \sum_{\langle i,j \rangle} \aopd_i \aop_j
    + \frac{U_0}{2}\sum_{i=1}^3 \nop_i \left(\nop_i-1\right)
    + U_1 (\nop_1 \nop_2 + \nop_2 \nop_3) + U_2 \nop_1 \nop_3,
\end{eqnarray}
where the number operators are defined as $\nop_i = \aopd_i \aop_i$ and $\langle i,j \rangle$ sums over neighbouring sites. The hopping rate is given by
\begin{eqnarray}
J=-\intvol \phi_i^*(\vecr) \bb{-\frac{1}{2}\laplace + V(\vecr)} \phi_{i+1}(\vecr)
\end{eqnarray}
and the interaction is parametrised by the three parameters
\begin{eqnarray}
U_k = \intvold \absvsq{\phi_i(\vecr)} \absvsq{\phi_{i+k}(\vecr')} 
\bb{4\pi a \, \delta(\vecr - \vecr') + \Vdd(\vecr - \vecr')}.
\end{eqnarray}
The on-site interaction $U_0$ includes parts of both dipolar and contact-interacting origin whereas the inter-site couplings $U_1, U_2$ only depend on the dipolar interaction, as the density-density overlap is negligible. For point-like, tightly localized wave functions $\phi_i$, the inter-site couplings satisfy $U_2=U_1/2^3$ as the dipolar interaction falls off like~$r^{-3}$. The following results, however, are also valid for extended wave functions and $U_2 = U_1/\alpha$ with $4\le\alpha\le8$.

In the special case of $J=0$ the model can be solved analytically and four distinct phases appear~\cite{Lahaye2010}. We quickly review them here to compare with our results. For $U_0 > 0$ and $U_1/\absv{U_0}\le8/15$ as well as for $U_0 < 0$ and $U_1/\absv{U_0}<-8$, the phase~\textbf{A} is present with
\begin{eqnarray}
n_1=n_3=\left\lfloor\frac{8(U_0 - U_1)}{24U_0-31U_1}N\right\rfloor
\end{eqnarray}
where $\lfloor\cdot\rfloor$ denotes the integer part. The other three phases are described by a single fixed ground state. Phase~$\textbf{B}$ appears for $U_0 >0$ and $8/15 \le U_1/\absv{U_0} \le 8$ and is characterized by the ground state $n_1=n_3=N/2$. For $U_0 > 0$ and $U_1/\absv{U_0} > 8$, as well as for $U_0 < 0$ and $U_1/\absv{U_0}>-1$, phase~\textbf{C} is present where all particles are occupying a single well. The central well is favoured if (weak) tunneling is present and thus we describe this phase by $n_2 = N$. The remaining part of parameter space $U_0 < 0$, $-8 < U_1/\absv{U_0} < -1$ is filled with phase~\textbf{D}, having two degenerate states with $n_1=n_2=N/2$ or $n_2=n_3=N/2$.

\subsection{Restrictions of the \bh model}
\figopt{0.7}{0.6}{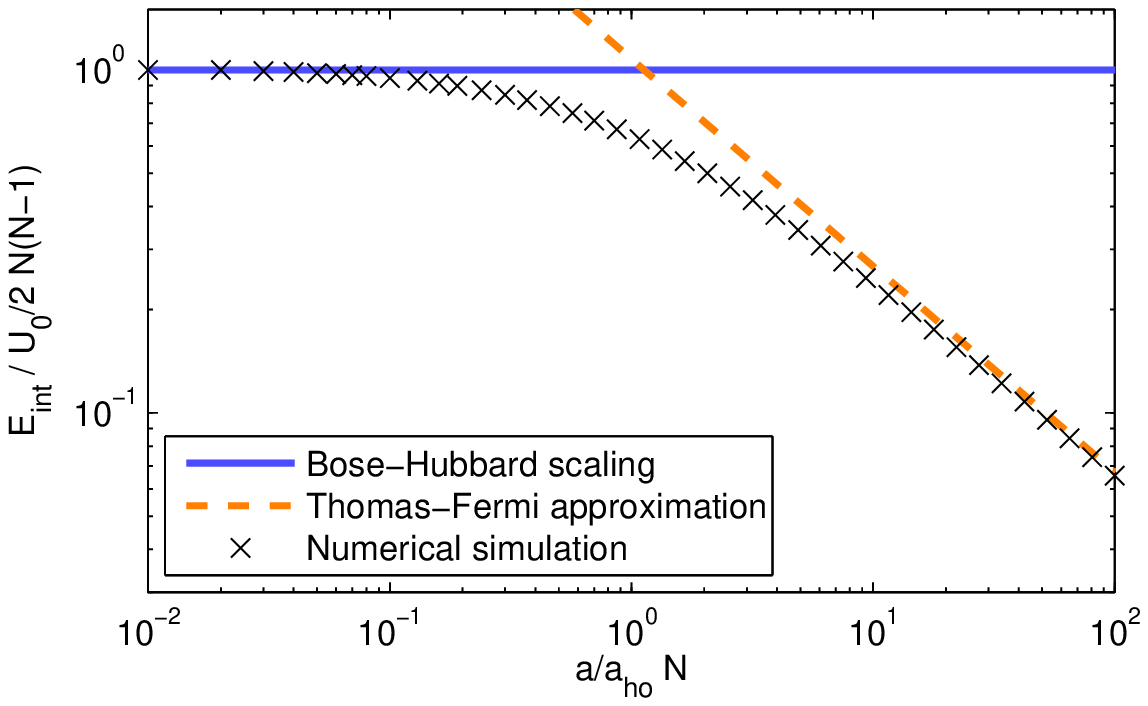}{Interaction energy of a contact interacting gas in a harmonic trap, compared to the energy term $U_0/2\cdot N(N-1)$ of the \bh model. The quadratic scaling is only reasonable for $a/\aho N \ll 1$. The strongly interacting regime is well described by the Thomas-Fermi approximation.}{b}

There are two main issues with the \bh approach that we will address in this section. Both restrictions arise from the assumption that a ground state wave function $\phi$ exists which does not depend on the number of particles and the interaction strength.

The \bh method intrinsically leads to a stable ground state solution as this is a premise of the model. This assumption, however, does not hold true in general for interacting quantum gases. As we would like to describe both repulsive and attractive interactions, this problem is of relevance in our system. The stability issue will be discussed in \sref{sec:stability}.

The second assumption is that the parameters $J$ and $U_k$, which are calculated by means of the single particle wave function $\phi$, are constant for all particle numbers $N$. This approximation is certainly good for small particle numbers and small values of $a, \add$. We demonstrate, however, that it is not well suited in our case.

For simplicity we consider a purely contact interacting \bec of $N$ particles in a spherically symmetric harmonic well with frequency $\omega$. We calculate the interaction energy as a function of $a/\aho N$, where $\aho = \sqrt{\hbar/m\omega}$ is the harmonic oscillator length. The \bh approach suggests a quadratic scaling $\Eint = U_0/2\, N(N-1)$ with the number of particles.

For $a/\aho N \ll 1$ this relation is a valid approximation, see \fref{fig:figure2}. However, we have in mind a system of at least $100$ atoms with a typical scattering length of $a \approx 5\unit{nm}$ and traps with a width of $\aho \approx 500\unit{nm}$. The resulting factor of $a/\aho N \approx 1$ is just in the crossover region of the diagram. For this value the interaction energy calculated by the quadratic term is already $35\%$ off, compared to the numerical simulation.
The extension to dipolar interacting gases is further increasing the problem. Inter-site repulsion or attraction can lead to changes of the neighbouring on-site wave functions.

\subsection{Mean-field approach}
For reasons being apparent now, we will use an alternative approach and describe the system in a mean-field picture. We stress that this approach requires, contrarily to the \bh model, that the particle number $N\gg 1$. The \gpe for our system is given by~\cite{G'oral2000}
\begin{eqnarray}
    \label{eq:dimgpe}
        i\partiald{t}\Psi(\vecr,t) = \left[
            -\frac{1}{2} \laplace
            + V(\vecr) 
            + 4\pi a (N-1) \left|\Psi\right|^2
            + \Phidd(\vecr,t)
        \right] \Psi(\vecr,t)
\end{eqnarray}
where we have introduced the condensate wave function~$\Psi(\vecr,t)$ which we normalize to unity. The dipolar interactions are included by the mean-field potential
\begin{eqnarray}
    \Phidd(\vecr,t) &= 3 \add N \intvolp \frac{1 - 3 \cos^2(\vartheta)}{\absv{\vecr - \vecr'}^3} \absvsq{\Psi(\vecr',t)}.
\end{eqnarray}
We find the ground state of equation~\eref{eq:dimgpe} by imaginary time evolution on a 3D grid~\cite{Chiofalo2000}. The dipolar interaction part is efficiently computed in momentum space by means of fast Fourier transformations, as the mean field potential $\Phidd$ has the form of a convolution.

Although both the \bh Hamiltonian and the \gpe are derived from the same multi-particle Hamiltonian in second quantized form there is no direct link between the two models. This implies that there can be no relation between $a, \add, N$ and the \bh parameters $U_0, U_1$ since the latter depend on~$\Psi$ which is not fixed in the mean-field approach.

\subsection{Stability} \label{sec:stability}

Dipolar quantum gases have a complex stability behaviour~\cite{Koch2008} which leads to some peculiarities when treating them numerically. Typically, a critical scattering length~$\acrit$ can be defined, which depends on the geometry of the external potential and the strength of the dipolar interaction~\cite{Koch2008}. For all scattering lengths $a<\acrit$ the system is unstable and no ground state can be found. The crossing of the stability threshold leads to a collapse of the condensate wave function which can easily be identified in the simulation. The collapse in a single harmonic trap has been studied in detail~\cite{Koch2008}. Above the critical value of the scattering length, a stable solution can be found for any $a>\acrit$.

As we are going to trace out the stability threshold of the \tw configuration we need to assure that the simulation is not crossing any unstable regions during the imaginary time evolution. We proceed as follows: The simulation is set up with an initial Gaussian wave function which spreads over all wells. The spreading is such that the width in $z$-direction is equal to the spacing between the wells to assure a certain fraction of particles in each well. We stress that the final state does not depend on the chosen initial wave function. The Gaussian can even be placed asymmetrically over one of the outer wells and the imaginary time evolution still yields the same (symmetric) ground state.

In the first sequence of imaginary time evolution all interactions are set to zero ($a=\add=0$) and we reach the ground state for the non-interacting case, see \fref{fig:figure3} for $\add=0$. This is our starting point to reach any point in the parameter space. The diagrams in figures~\ref{fig:repulsive} and \ref{fig:attractive_combined} are scanned line by line from right to left. We start at a high scattering length $a$ with $\add$ still set to zero to assure that we are in the stable region. After the ground state for the contact interacting case is reached we select the final value of $\add$ and probe one horizontal line in the diagram by subsequent runs of imaginary time evolution while decreasing $a$ until the wave function collapses to basically one grid point. At this point the simulation has reached the critical scattering length $\acrit$.

\subsection{Geometry}
Performing the numerical simulations, we have to choose a fixed geometry and strength for the external potential, like the widths of a single well in all spacial directions $w_x, w_y, w_z$ and the depth of the potential $V_0$. We reduce the parameter space by choosing reasonable values for the parameters, having symmetries as well as physical limitations in mind.

The width in $z$-direction is restricted, as the different wells are not clearly distinct for $w_z \gg 1/2$ (remember that we are measuring lengths in units of the spacing between two wells). Oppositely, for $w_z \ll 1/2$ the tunneling is too low to reach the ground state of the system with imaginary time evolution (or in an experiment). In the simulations we will therefore set $w_z$ to a value of $1/2$.
We ask for two conditions to fix the values for the two remaining widths $w_x$ and $w_y$.
As we focus especially on inter-site effects, changing the polarization direction (from an attractive inter-site coupling to a repulsive one) should not change the on-site effects~\footnote{There might be changes in the on-site energy due to second order effects: if the changed inter-site coupling leads to a different shape of the on-site wave function. Strictly speaking, this condition can only be fulfilled for a single well.}. To satisfy this requirement, the width in one of the directions perpendicular to $z$ has to be equal to $w_z$. Without loss of generality we define $x$ to be the polarization direction for the \qu{repulsive geometry} ($z$ for the \qu{attractive} case). Therefore we need to set $w_x = w_z$.

The second condition concerns the stability. To see a large variety of states we want the stability of a single well to be higher than in the spherical case (lower critical scattering length $\acrit$ for the same $\add$). To fulfil this, the remaining width $w_y$ has to be larger than the two others~\cite{Eberlein2005}. This leads us to cigar-shaped traps which are placed side by side, as shown in \fref{fig:pot_geometry} \hyperref[fig:geometry_rep]{(b)} and \hyperref[fig:geometry_attr]{(c)}. We fix the trap aspect ratio to a value of $\omega_z/\omega_y = 1/8$ as this turns out to be a reasonable value for an experiment, too. We stress that simulations with different aspect ratios $\omega_z/\omega_y<1$ do not show a qualitatively different behaviour.

For the depth of the potential there are also certain limitations. If $V_0$ is too low, the potential is not able to trap the particles. If it is too large, the tunneling rate is suppressed (see above). It turns out that $V_0 = 80$ is a reasonable value which allows for a large diversity of ground states. Again, additional simulations show that the behaviour is not sensitive to the precise value of this parameter, even quantitatively.

\subsection{Interaction}
We simulate the dimensionless \gpe~\eref{eq:dimgpe}. As all parameters of the external
potential are fixed, there are only two free quantities. These are the values of the contact and dipolar interaction strength given by the
dimensionless products $a N$ and $\add N$. Note that it is not necessary to change the number of particles $N$ independently. 
In the following we will present simulations where we change both $a N$ and $\add N$ to scan the remaining parameter space.
Note also that in our dimensionless units the values of $a$ and $\add$ depend on the spacing between two lattice sites.

\section{Results\label{sec:results}}
\figopt{0.8}{0.6}{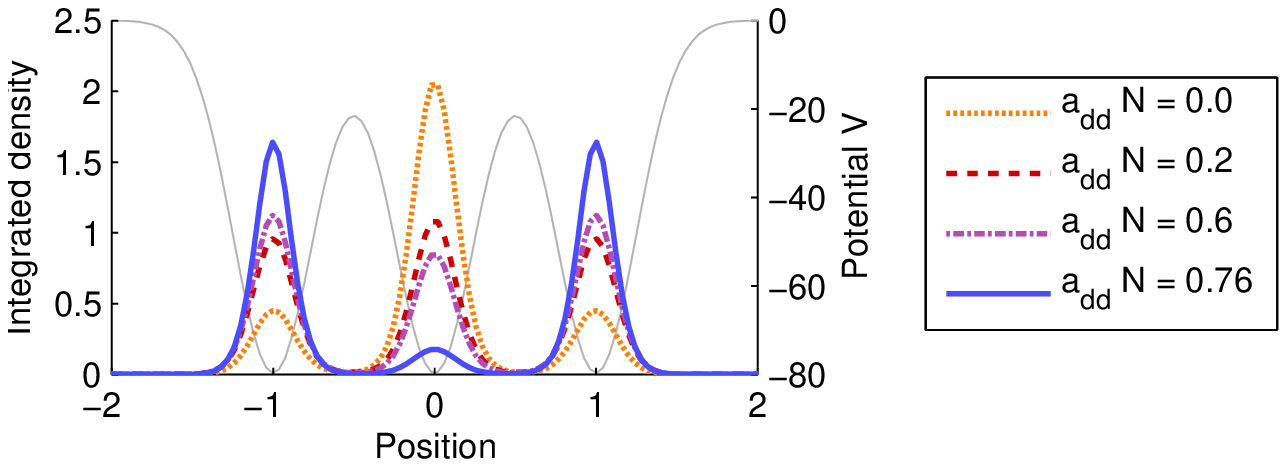}{Purely dipolar states for $a = 0$ and different values of $\add N$ in the repulsive case. Particles move to the outer wells as the dipolar inter-site repulsion grows.}{b}

To analyze the structure of the states we plot the ratio $n = (n_1+n_3)/N$ in analogy to~\cite{Lahaye2010}. In the simulation we calculate the occupation numbers $n_i$ by integrating the density $n(\vecr)$ over the volume of the $i$-th well. We have divided the whole volume of the simulation into three parts such that $\sum n_i=N$.

As we have a finite tunneling rate due to the finite potential depth and spacing, the states found with the mean-field calculations are always symmetric ($n_1 = n_3$) with respect to the central well. The two asymmetric states in the \textbf{D} phase found in the \bh approach are only present for tunneling $J=0$. For $J>0$ the symmetric and anti-symmetric combination of both states split in energy and yield a symmetric density distribution.

Let us first discuss the non-interacting ground state of the \tw system. In a simple 3-mode approach we can use localized wave functions $\Psi_i = \sqrt{n_i/N} \phi_i$, centred at the $i$-th well, as defined in \sref{sec:bhmodel}. If the ground state energy of a single well is $E_0$ and the overlap integral for neighbouring wells is $J$, we have to diagonalise 
\begin{eqnarray}
H=\begin{pmatrix}
E_0 & -J & 0 \\
-J & E_0 & -J \\
0 & -J & E_0
\end{pmatrix},
\end{eqnarray}
from which we immediately find the ground state $(1/2, 1/\sqrt{2}, 1/2)$ with occupation numbers $n_2 = N/2$ and $n_1=n_3 = N/4$, giving a ratio of $n=1/2$. Close to the origin of the diagrams in \fref{fig:repulsive} we find indeed states with $n\approx 1/2$ (see also \fref{fig:figure3} for $\add N=0$). Note that the non-interacting ground state in the \bh model for $U_k=0$ is given by 
\begin{eqnarray*}
\bb{{\aopd_1}/{\sqrt{2}} + {\aopd_2}/{2} + {\aopd_3}/{\sqrt{2}}}^N\ket{000}
\end{eqnarray*}
which also yields $\langle n_2\rangle=N/2$ and $\langle n_1\rangle=\langle n_3\rangle=N/4$.

Adding a repulsive contact interaction leads to a flattening of the density profile in the sense that we expect to have a uniform distribution $n_1=n_2=n_3=N/3$ (or a ratio of $n=2/3$) for large scattering lengths. For $\add=0$ and $a>0$ we observe states with $1/2 \le n \le 2/3$. Every state that is observed for $\add > 0$ which has a ratio $n$ outside this interval is thus a clear indication of the dipolar inter-site effects.

\subsection{Repulsive inter-site interactions}

\begin{figure}[t]
    \begin{center}
	    \subfigure[repulsive case]{
		    \ig{0.3}{0.4}{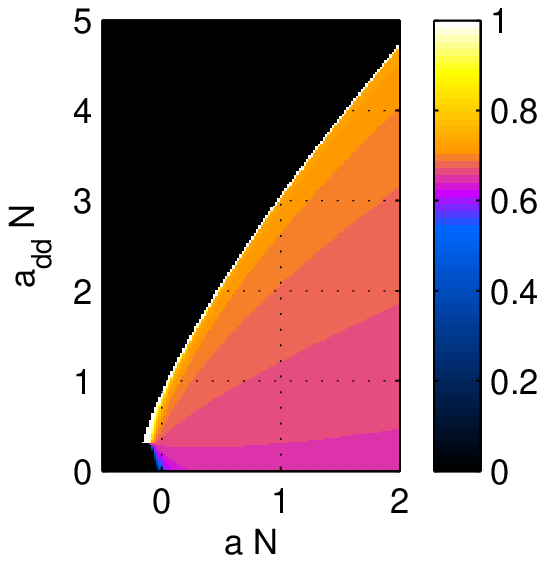}
		    \label{fig:cigars_repulsive}
	    }
	    \subfigure[horizontal cuts]{
		    \ig{0.517}{0.4}{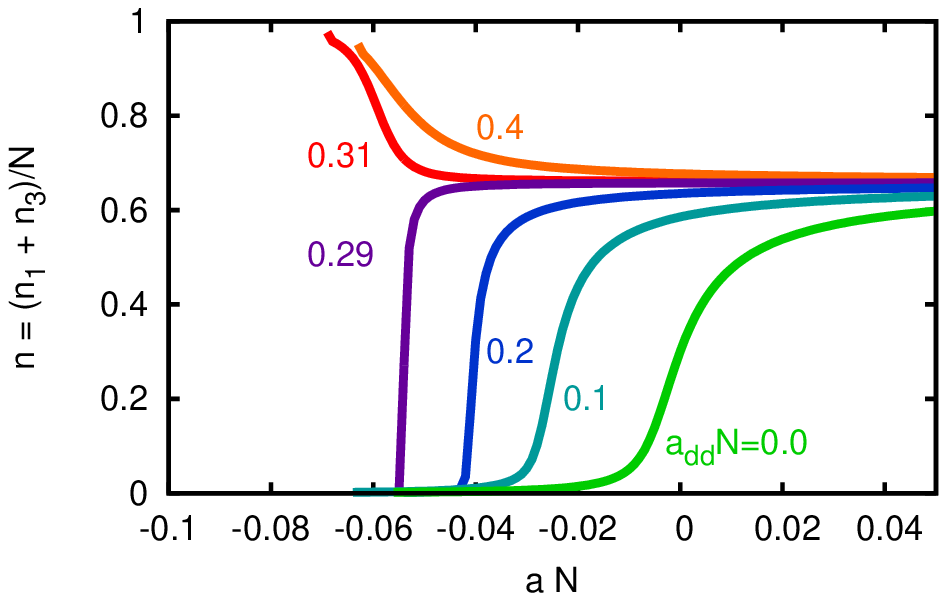}
		    \label{fig:ratio_cuts}
	    }
	\end{center}
	\caption{The quantity $n = (n_1 + n_3)/N$ is plotted for the geometry with repulsive inter-site interactions, see \fref{fig:geometry_rep}. (a) Black coloured areas indicate regions of instability where no ground state could be found. For the purely contact interacting case ($\add = 0$, $a > 0$) we find an equally populated state with $n_1 \approx n_2 \approx n_3$. For increasing $\add$ the state transforms into the state with $n_2 = 0, n_1 = n_3 = N/2$ (white). This state only appears close to the instability border. (b)~Horizontal cuts through the phase diagram on the left are shown for different values of $\add N$, as indicated by the labels. The behaviour changes qualitatively around the threshold value $\add N \approx 0.3$. Note that only a narrow region of the  phase diagram is shown in the cuts.  \label{fig:repulsive}}
\end{figure}

\Fref{fig:repulsive} shows an overview of the states found by imaginary time evolution for the geometry with repulsive inter-site interactions. We find the whole spectrum $0 < n < 1$. Once $n$ approaches the value of $0$ or $1$, the states become unstable. In particular, we observe ground states with a ratio of $n>2/3$, implying that there are fewer particles in the middle well than in the outer ones ($n_2<n_1=n_3$), a clear indication of the inter-site repulsion. States with $n \approx 0$ appear even in the purely contact interacting case for $\add N=0$, see \Fref{fig:ratio_cuts}, and are therefore less suited to demonstrate the long-range nature of the interaction. The on-site attraction for negative $a$ is enough to concentrate the atoms in the central well until the condensate finally collapses for $a N \approx -0.055$.

\Fref{fig:ratio_cuts} also reveals that a threshold value exists at $\add N \approx 0.3$ with a sudden change of behaviour. For dipolar interactions weaker than this critical value, the ratio $n$ is always lower than $2/3$. This region is dominated by the on-site interactions. For decreasing contact interaction the ratio smoothly approaches $n \approx 0$ and finally collapses. Contrarily, for an interaction strength larger than $\add N\approx 0.3$, the ratio $n$ increases when lowering the scattering length until it finally reaches $n\approx 1$ and collapses.
The behaviour of the border, separating stable from unstable regions, is also different below and above the threshold. Below, the instability is triggered by a collapse in the middle well as it holds most of the particles. The system gets more stable for higher values of $\add N$ (critical scattering length decreases with growing dipolar interaction).
Above the threshold, the instability is caused by the particle flow to the outer wells which finally leads to a collapse in either the left or the right well. This is a clear signature of the inter-site effects. In this regime, the system gets less stable for growing dipolar strength ($\acrit$ increases for growing dipolar interaction).
We remark that the threshold value is not depending on the aspect ratio of the single wells but changes with the depth of the potential.

\subsection{Attractive inter-site energy}
\begin{figure}[t]
    \begin{center}
	    \subfigure[attractive case]{
		    \ig{0.3}{0.4}{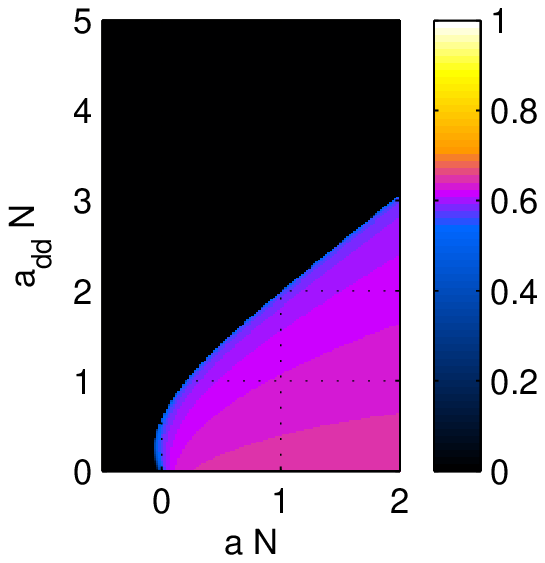}
		    \label{fig:cigars_attractive}
	    }
	    \subfigure[combined plot]{
		    \ig{0.3}{0.4}{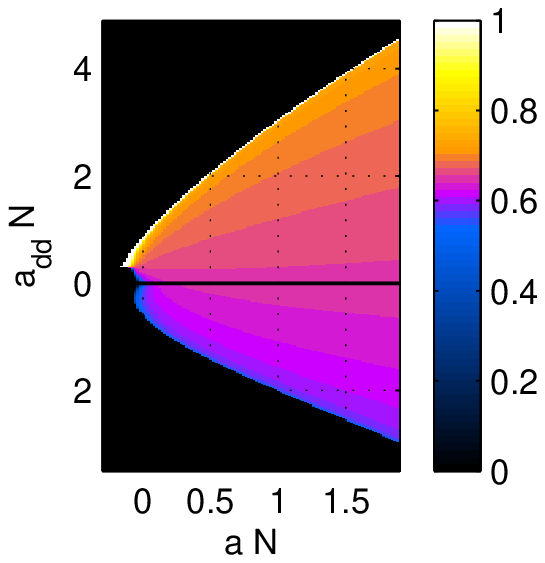}
		    \label{fig:cigars_combined}
	    }
    \end{center}
	\caption{(a) The quantity $n = (n_1 + n_3)/N$ is plotted for the geometry with attractive inter-site interactions, see \fref{fig:geometry_attr}. Black colour again indicates unstable regions. Close to the instability, states with $n \approx 0.5$ form which resemble the non-interacting ground state. (b) Combined plot which shows the repulsive case (upper part) and the attractive case (lower part, flipped $\add$ axis). Both situations match exactly if $\add = 0$ but also for larger $\add$ the plot combines to a consistent picture. Differences in both parts are solely caused by inter-site effects. \label{fig:attractive_combined}}
\end{figure}

Figure~\ref{fig:cigars_attractive} shows an overview of the states for the geometry with attractive inter-site interactions. We can immediately see that the inter-site attraction is destabilizing the system as it has a larger $\acrit$ for most values of $\add$.
We observe the whole spectrum of ratios $n$ between the equally populated state with $n = 2/3$ and $n = 0$. Approaching the value of $n = 0$, the states become unstable. In this situation, the collapse is again initiated in the middle well. As we do not observe any states with $n$ outside the range $0 \le n \le 2/3$, the attractive geometry is not suited to demonstrate the inter-site effects doubtlessly.

\subsection{Combined picture}
\Fref{fig:cigars_combined} shows a combined plot of both cases (repulsive and attractive inter-site interaction). For $\add = 0$ both situations are identical. Differences in the upper and lower part are solely caused by inter-site effects, as the geometry of the \tw potential was designed in such a way (the on-site energy does not change when rotating the polarization direction).

We now compare our findings to the results of the \bh approach.
Within the mean-field theory and our numerical simulations we observe the counterparts of
all states of phase~\textbf{A}, as well as states close to those of the phases \textbf{B} and \textbf{C}. For phase \textbf{D} with its two degenerate states $n_1=n_2=N/2$ and $n_2=n_3=N/2$, the comparison is a bit subtle. The mean ratio of both states is $n=1/2$, which can be seen in the simulations. However, these states are related to the non-interacting state which also has $n=1/2$. The mean-field approach is unable to distinguish both cases and finds only the symmetric states.

As we do not observe extended regions in the phase diagram with $n=0$ or $n=1$ we conclude that the regions which correspond to the phases \textbf{B}, \textbf{C} and \textbf{D} are unstable, an aspect which is not observable in the \bh approach. We remark, however, that the presented theory is only valid for large particle numbers. For small samples with few particles, the \bh approach is well justified and stable phases should appear.

Even in the absence of extended phases, the clear indication of inter-site effects is still visible. Especially the ground states with $n > 2/3$ above the threshold value at $\add N \approx 0.3$ do not appear for purely contact interacting condensates and should therefore be considered as a strong evidence for dipolar inter-site interactions.

Finally, we justify the range of values used for the parameters $a N$ and $\add N$. We recall that the unit of length we use is the spacing between two wells. We adopt the suggested experimental parameters of~\cite{Lahaye2010} where a spacing of $l=1.7\unit{\upmu m}$ is used. For $2000$ $^{52}\text{Cr}$ atoms with a dipolar length of $\add \approx 0.79\unit{nm}$, the (dimensionless) value of $\add N \approx 0.93$ has the right order of magnitude. The scattering length $a \approx 5.8\unit{nm}$ leads to the dimensionless value $a N \approx 6.8$. This value, however, can be tuned precisely by means of a Feshbach resonance~\cite{Lahaye2007}, thus allowing for the detection of the interesting states close to the border of instability. Smaller samples of atoms could still provide the necessary dipolar interaction strength when using different species like Dy with a dipolar length of $\add\approx 7.1\unit{nm}$~\cite{Lu2011}.

\section{Conclusions}
In this paper we studied the properties of a dipolar quantum gas, loaded into a \tw potential.
Using the numerical solution of the nonlocal \gpe, the phase diagram of the system has been obtained. In particular, we identified the range of parameters where the ground states reveal strong inter-site effects and we traced out the instable regions in the phase diagram.
We find that ultracold gases of atoms with a high magnetic moment, like Cr or Dy, are suited to demonstrate these features.
The presented ground states were compared to the results of the dipolar \bh model.

\ack \label{sec:acknowl}

We thank M. Jona-Lasinio and L. Santos for useful discussions. Two of us (K. P. and K. Rz.) acknowledge the support from the (Polish) National Science Center
under contract No. DEC-2011/01/B/ST2/04069.
All authors acknowledge the financial support from the project \qu{Decoherence in long range interacting
quantum systems and devices} supported by contract research \qu{Internationale Spitzenforschung II} of the Baden-W\"urttemberg Stiftung and the support from the DFG (SFB TRR/21).

\section*{References}

\providecommand{\newblock}{}


\begin{thebibliography}{10}
\expandafter\ifx\csname url\endcsname\relax
  \def\url#1{{\tt #1}}\fi
\expandafter\ifx\csname urlprefix\endcsname\relax\def\urlprefix{URL }\fi
\providecommand{\eprint}[2][]{\url{#2}}

\bibitem{fisher1989}
Fisher M~P~A, Weichman P~B, Grinstein G and Fisher D~S 1989 {\em Phys. Rev.
  B\/} {\bf 40} 546--570

\bibitem{Jaksch1998}
Jaksch D, Bruder C, Cirac J~I, Gardiner C~W and Zoller P 1998 {\em Phys. Rev.
  Lett.\/} {\bf 81} 3108--3111

\bibitem{Greiner2002}
Greiner M, Mandel O, Esslinger T, H\"ansch T~W and Bloch I 2002 {\em Nature\/}
  {\bf 415} 39--44

\bibitem{greiner2002revival}
Greiner M, Mandel O, H\"ansch T~W and Bloch I 2002 {\em Nature\/} {\bf 419}
  51--54

\bibitem{Bloch2008}
Bloch I, Dalibard J and Zwerger W 2008 {\em Rev. Mod. Phys.\/} {\bf 80}
  885--964

\bibitem{Bakr2009}
Bakr W~S, Gillen J~I, Peng A, Foelling S and Greiner M 2009 {\em Nature\/} {\bf
  462} 74--75

\bibitem{Sherson2010}
Sherson J~F, Weitenberg C, Endres M, Cheneau M, Bloch I and Kuhr S 2010 {\em
  Nature\/} {\bf 467} 68--72

\bibitem{Endres2011}
Cheneau M, Fukuhara T, Weitenberg C, Schauss P, Gross C, Mazza L, Banuls M~C,
  Pollet L, Bloch I and Kuhr S 2011 {\em Science\/} {\bf 334} 200--203

\bibitem{calarco2004}
Calarco T, Dorner U, Julienne P~S, Williams C~J and Zoller P 2004 {\em Phys.
  Rev. A\/} {\bf 70} 012306

\bibitem{G'oral2002}
G\'oral K, Santos L and Lewenstein M 2002 {\em Phys. Rev. Lett.\/} {\bf 88}
  170406

\bibitem{Griesmaier2005}
Griesmaier A, Werner J, Hensler S, Stuhler J and Pfau T 2005 {\em Phys. Rev.
  Lett.\/} {\bf 94} 160401

\bibitem{Beaufils2008}
Beaufils Q, Chicireanu R, Zanon T, Laburthe-Tolra B, Mar\'echal E, Vernac L,
  Keller J~C and Gorceix O 2008 {\em Phys. Rev. A\/} {\bf 77} 061601

\bibitem{Fattori2008}
Fattori M, Roati G, Deissler B, D'Errico C, Zaccanti M, Jona-Lasinio M, Santos
  L, Inguscio M and Modugno G 2008 {\em Phys. Rev. Lett.\/} {\bf 101} 190405

\bibitem{Mueller2011}
M\"uller S, Billy J, Henn E~A~L, Kadau H, Griesmaier A, Jona-Lasinio M, Santos
  L and Pfau T 2011 {\em Phys. Rev. A\/} {\bf 84} 053601

\bibitem{Giovanazzi2002}
Giovanazzi S, G\"orlitz A and Pfau T 2002 {\em Phys. Rev. Lett.\/} {\bf 89}
  130401

\bibitem{Smerzi1997}
Smerzi A, Fantoni S, Giovanazzi S and Shenoy S~R 1997 {\em Phys. Rev. Lett.\/}
  {\bf 79} 4950--4953

\bibitem{Albiez2005}
Albiez M, Gati R, F\"olling J, Hunsmann S, Cristiani M and Oberthaler M~K 2005
  {\em Phys. Rev. Lett.\/} {\bf 95} 010402

\bibitem{esteve2008}
Esteve J, Gross C, Weller A, Giovanazzi S and Oberthaler M~K 2008 {\em
  Nature\/} {\bf 455} 1216

\bibitem{Riedel2010}
Riedel F, B\"ohi P, Yun L, H\"ansch T~W, Sinatra A and Treutlein P 2010 {\em
  Nature\/} {\bf 464} 1170

\bibitem{Zibold2010}
Zibold T, Nicklas E, Gross C and Oberthaler M~K 2010 {\em Phys. Rev. Lett.\/}
  {\bf 105} 204101

\bibitem{Lahaye2010}
Lahaye T, Pfau T and Santos L 2010 {\em Phys. Rev. Lett.\/} {\bf 104} 170404

\bibitem{Damski2003}
Damski B, Santos L, Tiemann E, Lewenstein M, Kotochigova S, Julienne P and
  Zoller P 2003 {\em Phys. Rev. Lett.\/} {\bf 90} 110401

\bibitem{Sengupta2005}
Sengupta P, Pryadko L~P, Alet F, Troyer M and Schmid G 2005 {\em Phys. Rev.
  Lett.\/} {\bf 94} 207202

\bibitem{Lahaye2008}
Lahaye T, Metz J, Fr\"ohlich B, Koch T, Meister M, Griesmaier A, Pfau T, Saito
  H, Kawaguchi Y and Ueda M 2008 {\em Phys. Rev. Lett.\/} {\bf 101} 080401

\bibitem{Koch2008}
Koch T, Lahaye T, Metz J, Fr\"ohlich B, Griesmaier A and Pfau T 2008 {\em
  Nature Physics\/} {\bf 4} 218--222

\bibitem{G'oral2000}
G\'oral K, Rza\ifmmode \mbox{\c{}}\else \c{}\fi{}\ifmmode~\dot{z}\else
  \.{z}\fi{}ewski K and Pfau T 2000 {\em Phys. Rev. A\/} {\bf 61} 051601

\bibitem{Chiofalo2000}
Chiofalo M~L, Succi S and Tosi M~P 2000 {\em Phys. Rev. E\/} {\bf 62}
  7438--7444

\bibitem{Eberlein2005}
Eberlein C, Giovanazzi S and O'Dell D~H~J 2005 {\em Phys. Rev. A\/} {\bf 71}
  033618

\bibitem{Lahaye2007}
Lahaye T, Koch T, Fr\"ohlich B, Fattori M, Metz J, Griesmaier A, Giovanazzi S
  and Pfau T 2007 {\em Nature\/} {\bf 448} 672--675

\bibitem{Lu2011}
Lu M, Burdick N~Q, Youn S~H and Lev B~L 2011 {\em Phys. Rev. Lett.\/} {\bf 107}
  190401

\end{thebibliography}
\end{document}